# The development of ground-based Gamma-ray astronomy: a historical overview of the pioneering experiments

Razmik Mirzoyan

Max-Planck-Institute for Physics, Munich, Germany

**Abstract**

The ground-based technique for imaging atmospheric Cherenkov telescopes became a rapidly developing and powerful branch of science. Thanks to this technique, over 250 very high-energy gamma-ray sources of galactic and extragalactic origin have been discovered. Many fundamental questions of astrophysics, astro-particle physics, the physics of cosmic rays and cosmology are the focus of this technique. In the past 33 years since the discovery of the first gamma-ray source, the Crab Nebula, the discipline has made remarkable progress. Today, the technology boasts highly sensitive telescopes capable of detecting a point source 100 times fainter than the standard candle, the Crab Nebula, in 25 hours of observation. Further developments in this technology led to the Cherenkov Telescope Array (CTA), the next-generation large instrument. The sensitivity of CTA will be several times higher than that of the current best instruments. This article presents a brief history of ground-based very high energy gamma-ray astrophysics.

## 1. Introduction

Jelly's classic book (Jelley, 1958) covers most aspects of bluish radiation and, despite its advanced age, is still a reference book for many researchers. A number of highly regarded books and articles were devoted to the history of Cherenkov emission and ground-based very high energy (VHE) gamma astrophysics (Jelley, 1982), (Lidvansky, 2005), (Watson, 2011), (Weekes, 2003), (Fegan, 2012), (Lorenz & Wagner, 2012), (Hillas, 2013). The interested reader can find more details in the recent very interesting book (Fegan, 2019) and in the articles from this author (Mirzoyan, 2013), (Mirzoyan, 2014). In this report, the author would like to take a personal look at key developments in ground-based VHE astrophysics, from its inception to the present day that have led to today's success.

## 1.1 The Very beginning

Oliver Heaviside was the first to calculate the motion of an electron in a transparent medium with superluminal speed. He discovered that such motion will be accompanied by a specific radiation (Heaviside, 1889). He published several papers on what at the time seemed an abstract problem. Unfortunately, they remained unnoticed for at least half a century.

Arnold Sommerfeld studied the problem of a charge moving in vacuum with superluminal speed. In a medium with a given refraction index *n* his equations provided a valid solution. He came to a similar conclusion that a special radiation shall accompany the motion of the charge (Sommerfeld, 1904).

Already in ~1910, Marie Curie observed in her dark cellar that flasks filled with solvents of radioactive radium salts were glowing in blue. Thinking that this must be some kind of luminescence, she didn't follow the effect (Curie, 1941).

Mallet, a French researcher, was the first to systematically study the bluish emission. He published three papers in 1926-1929. He provided the external description of the effect and even measured a continuous emission spectrum. He understood that the lack of emission lines and bands did not support the previously assumed origin of luminescence. It is a pity, but the key features of the bluish emission as its polarization and the anisotropy remained undiscovered by him (Malet, 1926), (Malet, 1928), (Malet, 1929).

## 1.2 Developments in 1930's

Pavel Cherenkov became a PhD student of Sergey Vavilov in 1930. He got the task to study the bluish "luminescence" emission. The sensors were his eyes (the classical photo multiplier tubes were not yet available), and for best sensitivity he had to spend long hours in the dark, cold basement of the institute.

Initially, he varied the temperature and pressure of radioactive liquids, but these did not affect the emission.

Also the special additives failed to quench the "luminescence". He discovered that light emission had a continuous spectrum. To his astonishment he found that even "pure" solvent liquids emitted light when bombarded by a radioactive emission. He wrote an article about his studies in 1934 (Cherenkov, 1934). His supervisor S. Vavilov declined to co-author that paper and instead wrote his own explanatory paper. Their articles appeared next to each other in the same issue of the journal (Vavilov, 1934).

It is interesting to recall how many effects are known that emit a continuous spectrum of light. This may allow one to trace the possible thoughts of S. Vavilov, who interpreted the continuous spectrum as electron bremsstrahlung.

Another three years later Cherenkov demonstrated in a simple, elegant experiment the anisotropic character of the emission. Motion of a charge was accompanied by light emitted along the surface of a conus of certain angular opening in the forward direction. His discovery paper got declined by the magazine *Nature* but later it was published in The Physical Review journal (Cherenkov, 1937). In the same year, theorists

Igor Tamm and Ilya Frank offered the explanation of that effect (Tamm, Frank, 1937).

Let us consider a dielectric medium with a refractive index *n* and the speed of electromagnetic interactions in it *c/n* (*c* is the speed of light in vacuum). A charged particle moving through such a medium polarizes it as it moves. In the case of a motion with the speed exceeding *c/n*, the polarization pattern cannot be symmetric with respect to the rapidly escaping particle. The relaxation of the polarized medium will be accompanied by emission of anisotropic radiation in the forward direction.

S. Vavilov suffered about ten heart attacks and eventually died in 1951 (while he was one of the top scientists in the hierarchy of the Academy of Sciences of the former Soviet Union, his famous geneticist brother died in prison, presumably from starvation). A. Cherenkov, I. Tamm and I. Frank were awarded the Nobel Prize in 1956.

### 1.3 Contribution of Cherenkov Emission from EAS into LoNS

Patrick Blackett studied the emission of light of the night sky (LoNS) and the aurora in 1948. He estimated that the Cherenkov radiation from elementary particles in extended air showers (EAS) accounts for $10^{-4}$ of the LoNS intensity (Blackett, 1948). During his visit to Harwell in 1952, he met young Jelly and Galbraith, who were experimenting with Cherenkov light emission in water. Blackett told them about his estimate of the Cherenkov light contribution to the LoNS intensity.

### 1.4 Discovery of Cherenkov emission in the atmosphere

Very short after that Galbraith and Jelly constructed the first air Cherenkov telescope. Their tiny telescope was based on a 25cm diameter parabolic mirror, fixed inside a dustbin and a single two inch PMT in its focus, see Fig.1.a). They pointed this simple telescope at the night sky and patiently waited. Soon they observed rare pulses coming from the sky. This discovery, published in 1953, laid the foundation of the atmospheric Cherenkov light detection technique (Galbraith, Jelley, 1953).

## 2. First generation atmospheric Cherenkov telescopes

### 2.1 Chudakov's telescopes in Crimea

Alexander Chudakov with colleagues built the first full-scale atmospheric Cherenkov telescope in Crimea, near the Black Sea shore in 1959 (Chudakov, et al.,1964). It was based on using 12 parabolic searchlight mirrors of F/D = 0.6m/1.55m, with a total area of 21 m$^2$, see Fig.1.b). Each mirror had a single PMT fixed in its focus. A diaphragm set in front of the PMT provided an angular aperture of 1.75°.

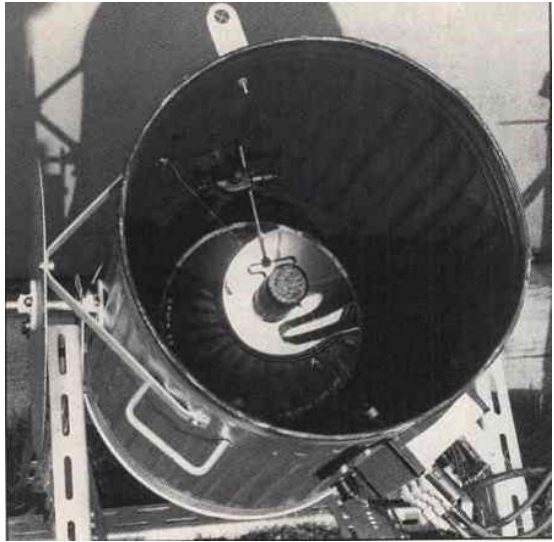

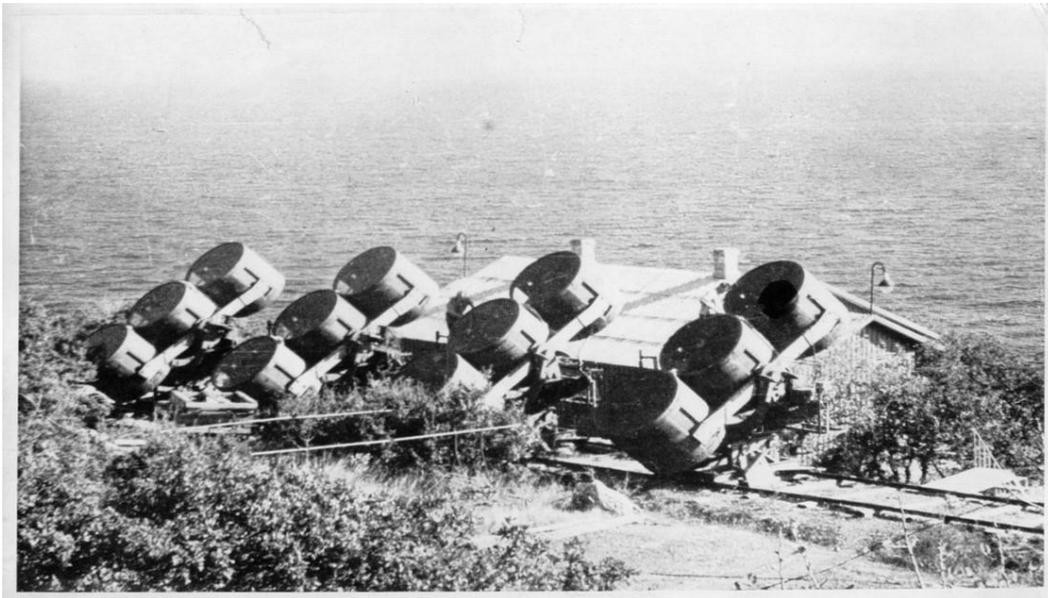

**Fig. 1** (**a**) The original detector of W. Galbraith and J.V. Jelley used in (Galbraith and Jelley 1953); photo courtesy of Trevor Weekes. (**b**) The telescope of A. Chudakov and colleagues in Crimea near the Black sea shore (Chudakov et al. 1964)

The choice of this value can be understood in an intuitive way; one can assume that the maximum emission angle of Cherenkov light is ~1° (here we ignore the multiple-scattering angle of electrons, the height dependence, etc.). Every three telescopes were fixed on a single mechanical mount. Four such mechanical mounts tracked the astronomical sources. A coincident system between them reduced the spurious triggers from LoNS and provided a lower threshold. Already then the crew members understood the importance of the different brightness of LoNS in On-source and the selected Off-source regions. To counteract this effect, they developed a simple circuit with feedback loop for providing constant illumination and stabilizing the rates. The pointing precision was 0.2° in elevation and 0.4° in azimuth.

The plan of Chudakov's crew was to measure a gamma-ray signal from a list of "prominent" celestial source candidates. To reduce the large aberrations of the mirrors and improve signal timing, Chudakov designed a lens of special form that was placed in front of the PMTs. The detection rate of EAS was about 3Hz and the threshold energy of gamma-rays was estimated to be ~3.4TeV. They used the drift-scan mode for observations. One points the telescope to an advanced in time position of the selected source and waits until it will pass through the field of view (FoV). By repeating this procedure many times, a significant amount of data can be collected overnight. The crew considered the radio sources as good candidates to observe (the X-ray sources and the pulsars had not yet been discovered). The list of observed sources included the Crab Nebula, Cygnus A, Cassiopea A, Virgo A, Perseus A, Sagittarius A. Also the clusters of galaxies Ursa Majoris II, Corona Borealis, Bootes, Coma Berenices. Even by today's standards, the list is certainly impressive. The experiment lasted 4 years. Unfortunately, no signal from any of the observed sources was measured, only flux upper limits were set. The upper limit from the Crab Nebula was about 20 times higher than its currently known flux. One can speculate that the gamma-rays from the Crab Nebula could have been discovered already ~60 years ago. But for achieving that goal the researchers should have observed it for several thousands of hours, i.e. exclusively and for the duration of many years.

G. Cocconi speculated at the international cosmic ray conference (ICRC) in Moscow in 1959 that a simple cosmic-ray instrument of a modest angular resolution (~1°), operating in the ~TeV energy range, will measure a very high gamma-ray signal to background ratio from the direction of the Crab Nebula (Cocconi, 1960). The results of Chudakov's crew showed that the speculations of Cocconi were overoptimistic. Interestingly, such speculations can spark curiosity and find followers who may ignite a chain of further developments.
Chudakov made a remarkable scientific career in his life and enjoyed a very high reputation on the international level. Until the discovery of the 9σ gamma-ray signal from the Crab Nebula by the Whipple team in 1988 he remained a great skeptic of the prospects of VHE gamma-ray astrophysics.
Chudakov's instrument used a large mirror area, important for achieving a low detection threshold, and a narrow time coincidence between separated by some distance telescopes. The latter suppresses the influence of LoNS on signal detection and also allows further lowering of the threshold.

## 2.2 Other first generation telescopes

In subsequent years, until the end of the 1980s, many other air Cherenkov experiments were built and operated aimed at measuring sources of VHE gamma radiation. With the exception of the 10-meter Whipple telescope, they were on a smaller scale than Chudakov's experiment. The lessons of Chudakov's experiment were well learned.
Typically, such experiments used a set of parabolic mirrors and fast PMTs at their foci, put into local coincidence. One example is the telescope built in the Glencullen Valley near Dublin. It used four 0.9 m diameter F/2 mirrors (total area ~2.5 m²) with fast PMTs in their foci, set into a coincidence scheme of 3.5ns gate width. This telescope was sent to Malta, where in early 1969 it began observations. As a result of observations, upper flux limits were established for several pulsars (Jennings, 1974).
In advanced configurations, the researchers placed the mirrors several tens of meters apart and organized a fast coincidence logic between them. Taking into account the practically flat transverse distribution of Cherenkov photons from TeV EAS on the ground up to the so-called "hump" of about 125 m (at an observation altitude of ~2 km a.s.l.), the largely spaced telescopes preferentially trigger on gamma-ray events. This suppresses hadrons due to their decreasing with distance lateral distribution of light, see, for example, (Fegan, 2019).

Some interesting, alternative designs are listed below.
Two telescopes of 6.5 m diameter reflectors were placed at a distance of ~120 m for stellar intensity interferometry in Narrabri, Australia. The researchers carried out observations of the Crab Nebula and two pulsars in 1968. No signal was measured (Hanbury Brown, Davis, Allen, 1969).
Grindlay and colleagues invented an observation technique, which they dubbed as "double beam". Each of their two telescopes had 2 PMTs. While tracking a selected source, one of the PMTs on each telescope was inclined towards the other one on 0.4° from the source direction; this allows one to better observe the shower maximum region. The other two PMTs were inclined to larger angles of 1.3° towards each other. The authors claimed that these allowed them to measure a signal from the so-called "muon core" of the showers and to suppress the hadron rate by two times (Grindlay, et al., 1975). Unsurprisingly, this claim sounds obscure and mysterious from today's perspective. Nevertheless, the principle was interesting; upgrades and modifications of it will be widely used in the future.
In fact, every PMT in an imaging camera is observing an individual solid angle in the sky.

The Haleakala telescope in Hawaii included six spherical coplanar glass mirrors of f/1 optics and a focus of 1.5 m, mounted on a single equatorial mount. Two independent groups of 18 PMTs in the focal planes observed different areas in the sky. The fast PMTs operated in the single photoelectron detection mode (Resvanis, et al., 1988). When several single photoelectron signals piled-up in a narrow coincidence window, a trigger was fired. Later it turned out that this detection method suffered from exhaustive rate of local muons.

The Nooitgedacht MK I telescope near Potchefstroom in South African Republic consisted of four mini-telescopes, placed on 55m from the central one. A mini-telescope contained three light detectors, consisting of 1.5m diameter, f/0.43 mirrors, focusing light on a PMT. MK II was the improved version of MK I telescope. It consisted of six mini-telescopes, set (225 – 322) m apart from each other. A single mini-telescope consisted of three mirrors, forming an f/1 optics of a focal length of 1.94m. A single PMT in its focus measured the On source region, whereas a ring-shaped mirror in the focal plane around the telescope's axis reflected the light from the 4.5° Off source region to a PMT installed close to the mirror level (a Cassegrain configuration). This allowed them to simultaneously measure the On and Off source regions. For details please see (Brink, et al., 1991) and the references therein.

The first generation telescopes of the Durham group were arranged similar to the logo sign of the Mercedes car, one in the middle and three in the corners of a triangle (Turver and Weekes, 1981).

The THEMISTOCLE array in south France followed a different approach. They used 18 widely spaced, tracking telescopes. Each telescope consisted of a parabolic mirror of 0.8m size and of a PMT in its focus. This array had a large collection area of $\geq 10^5$ m². It measured a gamma-ray signal from the Crab Nebula. The small area of individual mirrors and the large inter-telescope distances resulted in a relatively high energy threshold of ~ 3TeV and a low sensitivity (Baillon, et al., 1993).

PACT in India resembles the design to THEMISTOCLE, but has relatively high sensitivity due to the larger mirror area of individual stations and high number of stations (Singh, et al., 2009).

The HAGAR telescopes (Singh, et al. 2019) are located in Hanle in the Himalayas at 4500 m above sea level, on the same site as the 21 m diameter Cherenkov imaging telescope MACE, which is under commissioning (Borwankara, et al., 2020).

The seven telescopes of HAGAR, with a total mirror area of 31m², are placed at the center and corners of a hexagon of 50 m radius. Each telescope includes seven parabolic mirrors of 0.9 m diameter. Recently HAGAR researchers published results on the Crab Nebula detection (Singh, et al., 2019).

The AIROBICC instrument of HEGRA operated in 1992-2002 on the Canary island of La Palma. It consisted of 100 detector stations, measuring integrated Cherenkov light in a FoV of ~1 sr. Each detector was based on a single 8-inch PMT, coupled to a Winston cone-type light concentrator. The detectors were placed adjacent to HEGRA's particle (scintillator) detectors; The goal was to measure the ratio of Cherenkov light to charged particles from EAS and use it as a discriminant. The incoming direction of EAS was measured due to the arrival time differences of the shower front (Karle, Merck, Plaga, et al., 1995). Due to integration of LoNS in ~1 sr FoV the gamma-ray threshold of AIROBICC was estimated to be a few tens of TeV. The HEGRA array had a small size (~200m x 200m). This, combined with the high threshold and a marginal hadron suppression, did not provide sensitivity for measuring significant signals from sources.

Except for the 10m diameter Whipple telescope, which played a central role in giving birth to gamma astronomy (this will be discussed in some detail below), no major scientific results were achieved until the end of 1980s. On Fig.2 one can see a photo of the 10m diameter Whipple telescope.

The proceedings of the workshop series "Towards a Major Atmospheric Cherenkov Detector", can be recommended to readers interested in detailed information about those developments (see the list under the references).

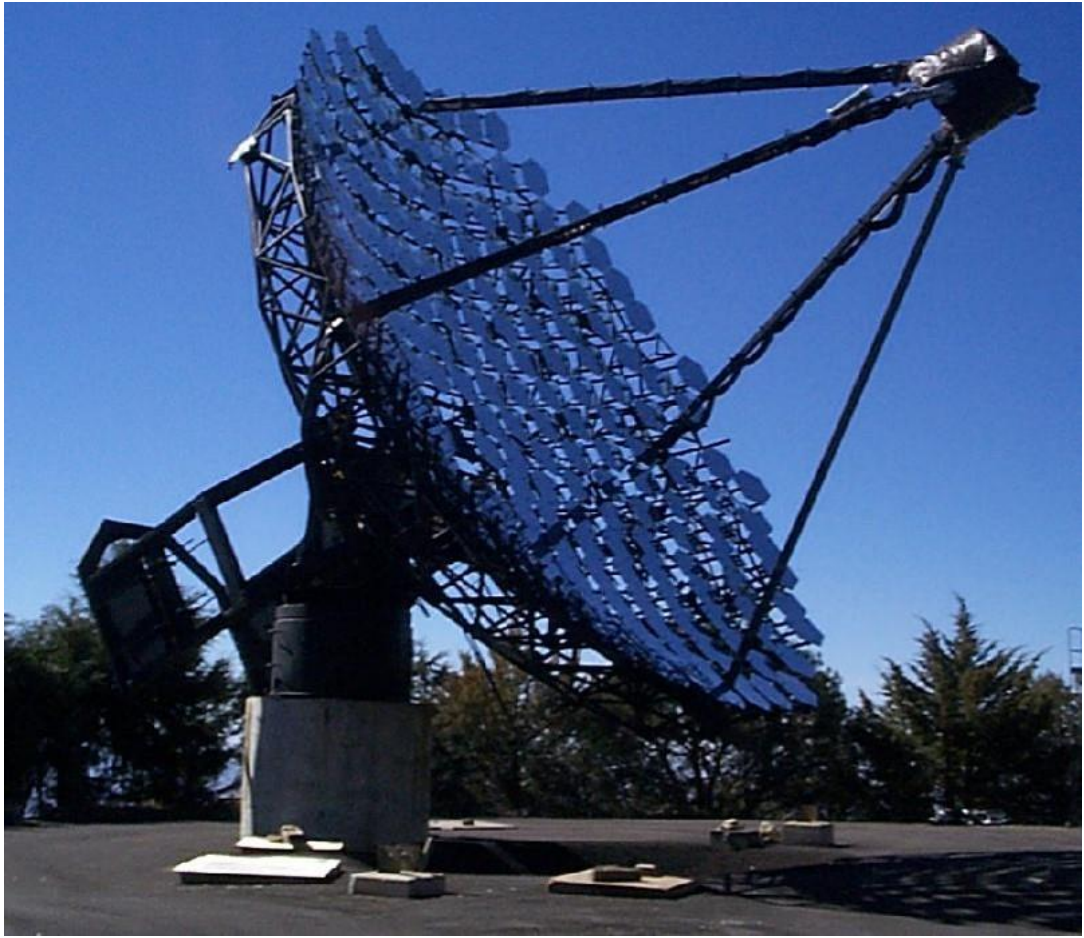

**Fig. 2** Photo of the 10 m diameter pioneering Whipple IACT on mount Hopkins in Arizona

## 2.3 A short summary on the first generation telescopes

The first generation experiments were based on counting the number of excess events from the ON source and selected OFF source regions. Fluctuations of LoNS have direct impact on PMT signals in telescopes. The instantaneous positive fluctuation of LoNS can add up with a genuine weak signal from a given shower and produce a trigger. It is natural that the ON- and selected OFF-source regions in the sky will have different LoNS intensity. These differences can cause a positive or negative excess in a counting rate experiment. As mentioned above in the text, researchers learned to counteract differences in brightness in the ON and OFF regions by setting a weak light source as for example, a miniature light bulb, near the input window of a PMT and built an electronic circuit to keep constant the sum light from the "bulb plus LoNS" system (Chudakov, et al., 1964). But whether the achieved accuracy was sufficient for long-term observations is another question. It should be kept in mind that those telescopes did not produce images, so it was not trivial to verify off-line the results of performed observations. The researchers understood the underlying problems and improved the technique. That paved the way for the next generation of telescopes.
Many first generation telescopes occasionally reported "signals" measured from a number of sources, even from pulsars. It can be assumed that sometimes some then unknown light phenomena in the upper layers of the atmosphere could cause a short-term sporadic excess. But in the case of pulsars, the likely cause was incorrect statistical and systematic data processing (see, for example, (Grindlay, 1972)). Later it became clear that those reports did not stand up to serious scrutiny. This can be understood when one considers the strong desire of small, enthusiastic groups of researchers to reinterpret small, minor, and/or sporadic excesses in observations as indicating the desired signal from sources. By then, the unspoken rule of "fashion" encouraged researchers to report source discoveries at conferences. By the end of the 1980s, the number of such source reports reached its apogee (see (Chadwick, McComb and Turver, 1990)). The impact of these vague reports from the early days proved important, as they underlined interest in the discipline and belief in the bright future of ground-based gamma-ray astronomy.

## 3 Image Shape of EAS

### 3.1 Air shower photos taken in Cherenkov light

The measurement of the shower image shapes by Hill and Porter in 1960 (Hill and Porter, 1960) is a real milestone in VHE gamma-ray astrophysics. They took fast shots of EAS images by using an image intensifier and a 25cm Schmidt telescope. At some point, these allowed one to realize the possibilities of ground-based gamma-ray astrophysics. Elliptical shower images were displaced from the direction of the source, and the shape of the image depended on the point of incidence of the shower axis. Hence, it is easy to guess that with the help of two telescopes separated by a certain distance, it would be possible to determine the arrival direction of the parent particle, as well as to suppress the background to a large extent. This important aspect was demonstrated about 30 years later by the HEGRA system of Air Cherenkov Imaging Telescopes. A similar goal was pursued by the Crimean telescopes GT-48, but their telescopes were separated by only ~20 m. Strong correlation of image parameters did not allow them to fully explore the benefits of observations with a distributed system of telescopes.

### 3.2 Monte Carlo simulations of EAS and the "stereo" observations

Viktor Zatsepin, one of the members in the Crimean experiment led by A. Chudakov, published in 1964 a remarkable research paper using the Monte Carlo method (Zatsepin, 1964). In it, one can find gamma-image isophotes common from today's point of view, as well as the angular distribution and the radial density of photons. The reading in this article is impressive, "since the maximum intensity of the shower light does not coincide with the direction of arrival of the primary particle, in studies in which the angular coordinates of the primary particle are determined by photographing the light flash from the shower, a greater accuracy of this determination should be achieved by photographing the shower simultaneously from several positions".

As early as about 60 years ago, some researchers clearly understood the potential of coincident (stereoscopic) measurements by using a distributed array of telescopes.

# 4 The second generation telescopes

## 4.1 The 10m Whipple telescope

In 1967, Giovanni Fazio and colleagues began construction of a 10 m diameter, F/0.7 telescope on Mount Hopkins at the Whipple Observatory (2300 m a.s.l.) (Fazio, et al., 1968a). As one can imagine, the large telescope provided a relatively low detection threshold for EAS. It began operating in 1968, at first with a single 5" PMT in focus, but later the number of PMTs was gradually increased to two and then to ten. These enabled simultaneous ON-source and OFF-source observations. In 1968, Fazio together with colleagues, including young Trevor Weekes, published a paper on the observations of 13 candidate gamma-ray sources. The Crab Nebula was prominently in the paper, along with M87, M82, IC443 (Fazio, et al., 1968b). Flux limits above the ~2 TeV threshold have been reported. Today we know that the listed candidates are very interesting sources of gamma-rays.

As early as 1977, T. Weekes and E. Turver proposed using a system of two telescopes 100 m apart, equipped with 37-pixel imaging cameras. The background could have been heavily suppressed (Weekes, Turver, 1977). The first imaging camera with 37 closely packed pixels, each with an aperture of 0.5°, was installed on the 10m Whipple telescope in 1983, see Fig.2.
Michael Hillas proposed at the ICRC in La Jolla in 1985 to parameterize the focal plane images by using the seconds moments of the measured charge distribution (Hillas, 1985). This turned out to be a real milestone in the development of the technique. Interestingly, this is one of the rare cases when a conference report received a huge number of citations and still continues to be cited.
Using this formalism, the Whipple team published in 1989 the result of the famous 9σ signal from the Crab Nebula (Weekes, et al., 1989).
This is considered as the birthday of the ground-based VHE gamma-ray astrophysics.
The scientific intuition and perseverance of Trevor Weekes and the small team around him paid off after about 20 years of effort and spawned a new branch of science.
A few years later, the 37-pixel camera with a 0.5°-pixel size was replaced by a finer resolution camera using pixels of 0.25°. This significantly improved the sensitivity of the telescope and lowered its threshold from ~700GeV down to ~300GeV.

## 4.2 GT-48 in Crimea

Since the late 1960's the group in the Crimean Astrophysical Laboratory (CrAO) led by Arnold Stepanian used two parabolic searchlight mirrors of 1.5 m diameter in coincidence for studying gamma sources. They reported detections of Cassiopea and Cyg-X3 in the beginning of the 1970's, especially the latter made a big resonance in the community. In the 1980's, the group started constructing a set of two large telescopes, separated by 20 m distance, named GT-48. On each mount they built six telescopes, three of the imaging type with 37 pixels and another three operating a single UV-sensitive, solar blind PMT. Every telescope had 4 mirrors of 1.2 m diameter and 5 m focus. The goal of the Crimean group was to profit from the stereo observations, see, for example (Kornienko, et al., 1993). They did not want to sacrifice neither the threshold nor the coincidence rate. Because of space limitation, they put the telescopes at 20m distance from each other. Their small mirror area and the low altitude of the location of 600 m a.s.l. provided a threshold of 900 GeV. The short distance between the telescopes did not allow them to fully exploit the differences in image parameters otherwise seen from largely separated detectors. In 1989, this installation was put into operation and in subsequent years it measured a number of sources.

### 4.3 High Energy Gamma Ray Astronomy (HEGRA)

The first HEGRA telescope was designed in 1990 as a somewhat modified version of the first Cherenkov telescope of the Yerevan Physics Institute (YerPhI) (Aharonian, et al., 1989). The latter was the prototype for the planned in 1985 "stereo" array of five telescopes. This array was later accepted by the HEGRA collaboration for construction. Its further developments became known in the community as the HEGRA imaging air Cherenkov telescope (IACT) array (Mirzoyan, et al., 1989).

The original plan was to build a distributed on 100m x 100m grid an array of 5 imaging telescopes, each 3m in diameter, with a tessellated reflector. The first telescope received a 37-pixel image camera in the 5m focal plane. The pixels used conical light guides (focons) made of UV-transparent Plexiglas and had an opening angle of 0.41° (Mirzoyan, et al., 1993). The imaging camera was based on PMTs with a GaP first dynode, offering incredibly high amplitude resolution. The mechanical mount of the first telescope was installed in late fall 1991 at the *El Roque de los Muchachos* observatory on La Palma while the imaging camera followed in mid-1992. The first signal from the Crab Nebula was measured after a couple of months of observations (Mirzoyan, et al., 1994). One year later the second telescope with the same pixel size but with one more ring in the camera (61 pixels), larger size reflector of 4.2m and alt-azimuth mechanical mount was built and put into operation at ~100m distance from the first one. Stereo observations began, the power of which was predicted in a special Monte Carlo study paper in 1993 (Aharonian, et al., 1993).

Four more telescopes were installed in the following years, of the same size as the second but with imaging cameras of 271-pixels, each of 0.24° in aperture, see Fig.3. Eventually, also the second telescope received a 271-pixel camera, and the array was completed in 1997. The last upgrade in the same year was the enlargement of the mirror surface area of the first telescope to 10.3 m².

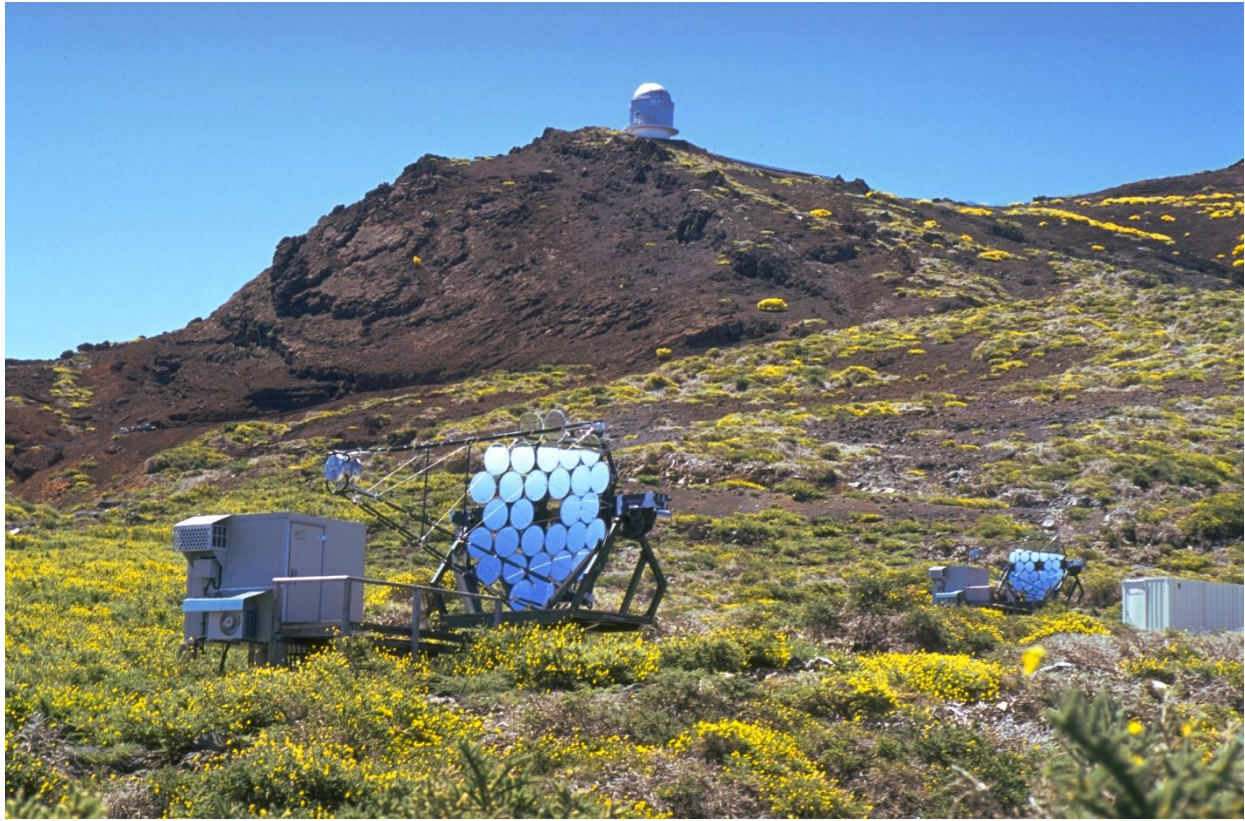

**Fig. 3** The observatory El Roque de los Muchachos on La Palma. Two of the six IACTs known as the HEGRA IACT array can be seen. The 2.5 m Nordic Optical Telescope can be seen in the center and top of the image

HEGRA operated until 2002. It demonstrated the advantages of stereo observations as the strong background suppression and hence the increased sensitivity and somewhat improved energy and angular resolutions, compared to a single telescope, see (Pühlhofer, et al., 2003) for details. HEGRA has produced a wealth of scientific results.

### 4.4 The Japanese 7-Telescope Array

The 7-Telescope Array was planned to include two big arrays of 127 imaging telescopes, operating in coincidence (Teshima, et al., 1992). A single telescope had a 3m diameter tessellated reflector and a 256-pixel imaging camera. In 1996-1997, three out of ready seven telescopes were installed in Dugway proving grounds, Utah, USA. The telescopes started taking data on the flaring MKN-501 and 1ES1959 in 1997. Unexpectedly, a missile launched by the military at the Dugway test-site lost its target position and mistakenly hit the telescopes' data acquisition containers. Luckily the rocket was unarmed and there was nobody in the test container, which of course was completely destroyed. This unusual event forced operations to cease at the end of 1997.

## 4.5  The CAT Telescope

The French CAT telescope was commissioned in late 1996 (Burrau, et al., 1998), on the same site as the previous non-imaging ASGAT (Goret, et al., 1993) and THEMISTOCLE (Baillon, 1993) telescopes. CAT was given a 600-pixel fine-resolution imaging camera of pixel size of 0.12°. Shortly after commissioning, the researchers found that due to the very fast pulses from the PMTs, the efficiency of detecting gamma rays was low. They slowed down the PMT pulses to ≥ 2.5 ns, thereby restoring high gamma-ray triggering efficiency. Also, the researchers installed several image cameras on the structure of the telescope. These helped to take into account the non-negligible bending of the telescope structure in the analysis of collected observational data. This was a successful telescope, which provided very interesting scientific results.

## 4.6  CANGAROO

CANGAROO was a collaboration of a number of Japanese Universities with the University of Adelaide. The collaboration started observing with a telescope of 3.8m size at a dark site 12 km north of Woomera, a desert town 500 km north of Adelaide, Australia. It used a single mirror of parabolic shape. The telescope was previously used for lunar ranging. Operations started in 1992. The location height was 165 m a.s.l. and the energy threshold was reported to be a few TeV. In the following years they discovered several new sources of gamma rays. A decade later, four telescopes of 10m size were built (CANGAROO-III). These made several discoveries and performed very interesting observations. Due to the on-going developments, the telescopes showed some differences in the design. It turned out that the collaboration was experiencing technical (first of all related to the targeted type of mirrors) as well as organizational problems, related to the data analysis. H.E.S.S. telescopes started their operation in 2002-2004. After the initial phase of observations, they could not confirm some of the CANGAROO results (Mori, 2007). A few years later this array terminated its operation.

## 4.7  Wide FoV Telescopes TACTIC and SHALON

The SHALON IACT in Tien Shan has a reflector of 10m² area and a 144-pixel imaging camera in the focus. It is claimed to be operational since 1993. The pixel size is 0.6°. The camera subtends a FoV of 7.2° x 7.2° in the sky. The author of this paper thinks that the used large pixel size plays a detrimental role for the telescope's sensitivity, setting a relatively high energy threshold despite the location height of 3338m

a.s.l. (Sinitsyna, et al., 1995). SHALON reported many results, mostly at various conferences, but unfortunately they did not stand up to critical discussion or comparison with any other known results.

A similar design has the TACTIC IACT on Mt. Abu in India, located at 1300m a.s.l. It is operating since 2001. TACTIC uses a reflector of 9.5m² area and a 6° FoV, 349-pixel imaging camera. Each pixel subtends an aperture of 0.31° in the sky. Only the central 3.4° x 3.4° FoV is set into the trigger logic (Koul, et al., 2007). Although TACTIC had a high energy threshold, likely due to sub-optimal light guide design, they reported several source detections.

### 4.8 The CLUE Telescope

The CLUE collaboration installed a distributed array of nine 1.8 meter telescopes at the HEGRA site on La Palma. To capture the images from EAS, they used a multi-wire, UV-light-sensitive proportional chamber (MWPC) in the focal plane. These had a composite pixel matrix of square electrodes in the back panel that were read out via capacitive coupling. The CLUE detector was filled with the admixture of TMAE gas. This substance should have a quantum efficiency of 5-15% when detecting Cherenkov light in the UV wavelength range 190-230nm. This experiment encountered technical problems, since, for example, the TMAE turned out to be a rather aggressive substance that interacted with the camera materials. There were also problems associated with the short distance of Cherenkov light transmission at the chosen wavelengths. Operating from 1997 to 1999, the telescopes reported positive observations of the well-known Crab Nebula, Mkn 421 and Mkn 501, and the lunar shadow (Bartoli, et al., 2001).

### 4.9 The Durham Mark 6 Telescope

The Durham telescopes went through many transformations in their design as well as changed their location where they performed observations. Some details one can find in (Chadwick, 2021). The imaging camera of the Durham Mark 6 telescope was based on 91 PMTs of one-inch size. These were surrounded by 18 PMTs of two-inch size. The telescope consisted of three dishes, where the central one was given the above described camera and the two side ones had crude 19-pixel cameras at their foci. A smart trigger logic was organized between the telescopes. One of the important results of this telescope was the discovery of PKS2155-304 (Chadwick, et al., 1999).

# 5   Solar power plants as gamma-ray telescopes

Up until the mid-1990s, it was believed that to lower the threshold energy of a given telescope by a factor of n, one would need to increase its mirror area by a factor of n².

It is interesting to follow this issue with the 1981 publication (Turver, Weekes, 1981). One can read there, "*The energy threshold of a simple detector is inversely proportional to the diameter of the light collector. An energy threshold of $10^{11}$ eV requires an effective aperture of 5-10 m. To get to $10^{10}$ eV requires an aperture of 50-100 m; such apertures would have been out of the question a few years ago but the development of large concentrators for solar energy research makes this energy threshold a realistic possibility*".

Note that in the early 1990s the threshold energy of what was then the largest 10m Whipple telescope with a reflecting surface of ~75m² was estimated to be ~300GeV.

Thus, to reduce its threshold by a factor of 10, to the range of 30 GeV, it was necessary to increase the area of its reflector by a factor of 100, i.e. mirror area 7500 m². And that was not realistic.

The problem was that due to the lack of a measuring instrument, the last unexplored decade on the electromagnetic scale, namely the energy range 10-300 GeV, was considered as "terra incognita". Many interesting physical phenomena were expected to occur there as due to lower absorption on extragalactic background light (EBL), the universe becomes progressively more transparent to gamma rays for lower energies. At the very low energies of ~20 GeV, signals from pulsars, from distant AGNs, from GRBs and from various transient events were anticipated.

The researchers were discussing about the possibilities if and how one can lower the threshold of a Cherenkov telescope by an order of magnitude.

Some enthusiastic researchers believed that the solar power plants, with their several thousand m² of mirror surface area, could offer a solution to achieve a very low energy threshold. Subsequently, several solar power plants were modified to work as gamma ray detectors. Four research groups were formed; STACEE (NM, US), CACTUS (Ca, US), CELESTE (France) and GRAAL (Germany and Spain). These took somewhat different approaches. For example, while the GRAAL team (Plaga, 2002) worked to collect reflected from heliostats Cherenkov photons into a single Winston cone of 1m entrance window, the STACEE team attempted organizing a modified imaging in the central light collection tower by directing light from individual heliostats to specific PMT channels (Jarvis, et al., 2010). For more details on converted into gamma-ray detector solar power plants, the reader is recommended to seek the very informative review (Smith, 2006). Some interesting studies have been made by using these modified arrays. CELESTE measured the flux from the Crab Nebula down to ~60 GeV (De Naurois, et al., 2002). Today's accurate measurements show that they underestimated the flux by 2.5 times.

## 5.1   The Solar power plants, the threshold energy and the MAGIC telescope

Operation of the MAGIC telescope proved that the above cited relation of the threshold on the mirror area was wrong.

The reason for this misconception is linked to the first generation telescopes. Unlike the non-imaging detectors, the fluctuations in the LoNS do not set the lower threshold for an imaging telescope. It gets "shared" among the large number of pixels in the camera and every pixel observes its own tiny solid angle in the sky. Additionally, these pixels are set into some coincidence scheme, which further suppresses the random fluctuations of LoNS.

A higher-level requirement is that a minimum amount of charge, on the order of ~100 photoelectrons, is needed to analyze an image and define its parameters (Mirzoyan, 1997a). The amount of charge scales linearly with the energy of an EAS, i.e., the threshold is inversely proportional to the mirror area (read the squared diameter of the reflector).

Realization of the latter relation played a key role for enabling successful operation of the IACT technique in sub-100 GeV energy range, down to 10-20 GeV. This was substantiated by proposing and building the pioneering 17m diameter MAGIC telescope project for sub-100 GeV gamma-ray astrophysics (Mirzoyan, 1997b).

As was predicted in the above cited two papers, the threshold is inversely proportional to the mirror area. An imaging telescope like MAGIC with ~240 m² mirror area could successfully perform measurement at energies above 40-50 GeV. It turned out that the "classic" imaging methods could offer a significantly higher efficiency than the detectors of the solar power plants, so that they stopped operating shortly afterwards.

# 6 The 3rd generation telescopes

The second-generation telescopes discovered only a handful of sources. However, it became clear that the "stereo" technology still had great potential, which was just waiting for wider use.

The third generation telescopes were designed before the potential of the second generation telescopes was fully exploited. Already in 1995 the first presentations on the concept of a 17m diameter IACT were made (Bradbury, et al., 1995),(Lorenz, et al., 1995). These were followed by the VERITAS letter of intent in fall 1996 and in the next year by H.E.S.S. Both VERITAS and H.E.S.S. were following the goal of doing astrophysics with a stereo system of 10 m diameter telescopes, based on well-known, proven technologies. These were well-known thanks to the Whipple telescope and the fresh experience with HEGRA. Instead, the design of MAGIC was aimed at performing sensitive measurements in the energy range below 100 GeV, down to 20-30 GeV, in "terra incognita". Obviously this task was significantly more demanding and challenging, several novel techniques and technologies were necessary for making it possible.
When HEGRA ceased its operation in 2002, the collaboration split into two parts. One part together with the scientists from Germany and France, formed the core of the H.E.S.S. collaboration and built their instrument in Namibia. The other part stayed in La Palma, at the original site of HEGRA, and the scientists from Germany, Spain and Italy founded the MAGIC collaboration.

The Veritas, H.E.S.S. and MAGIC telescopes have literally revolutionized astrophysics and made TeV measurements indispensable for the assessment and interpretation of the astrophysical sources. So many different types of sources have been discovered and studied in minute detail that it is impossible to give a compact description. The interested reader is advised to visit the (TeVCat) catalog for information and links to publications.

## 6.1 H.E.S.S.

The application of the H.E.S.S. collaboration was supported by the German and French financial agencies (while the VERITAS team had to wait for several more years to secure the financial support). The H.E.S.S. collaboration built their telescopes and started operation in Namibia in 2002-2004. Location of HESS in the southern hemisphere, very well suited for studying Galactic sources. Already in the beginning the H.E.S.S. team performed a scan along the galactic plane and made an impressive harvest of galactic sources. This array has turned out to be a very successful instrument, making a really high number of important discoveries and measurements above the energies (160-200)GeV, see, for example, (H.E.S.S. Collaboration, 2018). A very large telescope of 28m diameter has been set in the center of H.E.S.S. in 2012. This allowed them to perform observations also in the very low energy range of a several tens of GeV (Förster, 2014).
The reader can find more details on H.E.S.S. in the relevant chapter of this book.

## 6.2 VERITAS

The VERITAS telescopes, unlike H.E.S.S., who operate imaging cameras of ~5° geometrical apertures, use cameras of 3.5° FoV. Otherwise both instruments are similar and both have increased the originally planned 10 m diameter of their telescopes to 12 m. For some period, the exact location of these telescopes in Arizona remained uncertain. In the end VERITAS was built next to the administrative building of the Harvard-Smithsonian center for Astrophysics, not far from the foot of mount Hopkins and got inaugurated in 2007. As one would expect, also VERITAS turned out to be a very successful instrument that in the past years made a high number of important discoveries and measurements, see, for example (Mukherjee, 2018).
For more details, the reader is directed to the relevant chapter in this book.

## 6.3 MAGIC

MAGIC pioneered the technique to operate a ground-based VHE IACT in the energy domain from 300GeV, down to ~20GeV.
Initially, the 17m diameter MAGIC was proposed as a stand-alone telescope (Bradbury, et al., 1995),(Lorenz, et al., 1995). Innovative approach was necessary for operating such a telescope in the very low energy range, especially below 100 GeV, where a strong background from local muons was expected. The chosen strategy was to suppress the backgrounds by using an ultra-fast opto-electronic design of the telescope. These included a reflector of a parabolic design, specially developed very fast response PMTs, analog signal conversion into light and transport via optical fibers, a 2GSample/s FADC-based readout. These allowed the researchers to reduce the charge integration window down to only ~3 ns, thus effectively suppressing the contribution of LoNS.

The MAGIC-I telescope was built and put into operation in 2003-2004. For the first time MAGIC succeeded to significantly improve the sensitivity of an IACT, also due to the fast timing (Aliu, et al., 2009).

By developing a special SUM trigger configuration, the researchers succeeded operating the stand-alone MAGIC-I at a very low threshold of ≥ 25 GeV. This allowed them to detect for the first time a pulsed signal from the Crab pulsar (Aliu, et al., 2008). The second MAGIC telescope was set at 85m distance from the first one in 2009. The standard trigger of MAGIC made it possible to observe sources at energies above ~40-50GeV, see, for example, (Aleksić, et al., 2012).

The light-weight reflector frame from reinforced carbon fiber, coupled with the Active Mirror Control system, allowed to react promptly to alerts from satellite missions on transient sources like Gamma Ray Bursts (GRB). This feature paid-off in January 2019, in the first time allowing to detect the most intense VHE gamma-ray signal from GRB 190114C explosion, with only a delay of one minute (Acciari, et al., 2019),(MAGIC Collaboration, 2019).

The MAGIC telescopes introduced a number of innovations, some of which later became standard and were incorporated into Cherenkov Telescope Array (CTA) designs (Mirzoyan, 2022).

One of the main obstacles for achieving a low-threshold setting for an IACT is the adverse effect of after-pulsing in PMTs (Mirzoyan, et al., 1997c).

The trigger threshold of the two MAGIC telescopes was halved by using a so-called Sum-Trigger, which detects weak images in ~0.5° wide patches and includes a circuit to suppress the after-pulsing. Recently this new system allowed to detect a very weak signal from the Geminga pulsar at energies ≥ 15GeV (Acciari, et al., 2020).

# 7 The Fourth Generation Instruments

## 7.1 Cherenkov Telescope Array – The Major Instrument

The seven workshops "Towards a Major Atmospheric Cherenkov Detector", along with the first similar workshop that took place in 1989 in Crimea, former USSR (this could be considered as the workshop number "0"), and the last one in Palaiseau (France) in 2005, served their purpose (see the list under the references). In 2006 the researchers decided to unify the efforts of the entire community for building one major instrument, dubbed as CTA (Cherenkov Telescope Array). Today it counts over ~1500 researchers worldwide, working with the Atmospheric Cherenkov Technique. Advanced prototypes of the CTA telescopes were manufactured and tested, allowing the transition to the construction phase. Originally about 100 telescopes of 23m (Large Size Telescope - LST),

12m (Middle Size Telescope – MST) and 4m (Small Size Telescope – SST) sizes were planned to be built in the southern and northern observatories, covering the energy range from 10 GeV to more than 100 TeV (Acciari, et al., 2013), (The CTA Consortium, 2017). This is going

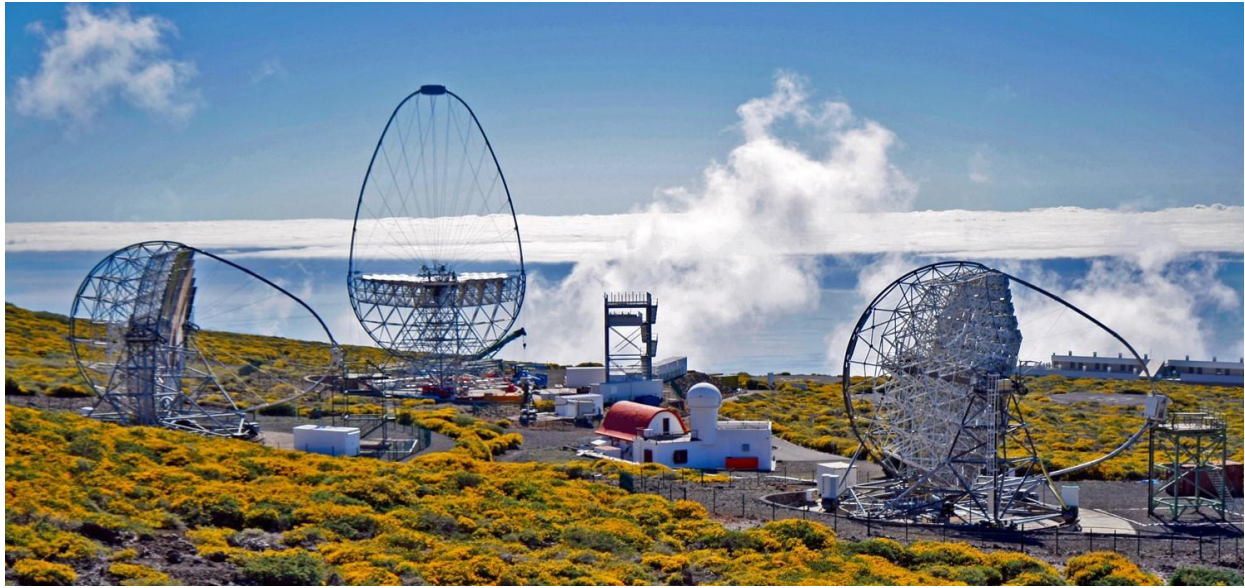

**Fig. 4** El Roque de los Muchahcos (ORM) observatory, 2200 m a.s.l., La Palma, Canary Islands. This is the location of the CTA North observatory. The CTA 23 m LST1 IACT is on the center-left, the 17 m MAGIC-I on the right, in-between the two the experimental house, the MAGIC-II IACTs on the left. One can see that the clouds over the Atlantic Ocean are lower than the observatory

to be the major ground-based instrument for doing astrophysics by means of gamma rays for the next few decades.

The first LST is in its final stage of commissioning at the ORM observatory in the Northern CTA location on the Canary island of La Palma, in close proximity to MAGICs, see Fig.4. LST-1 has already measured gamma-ray signals from dozens of sources. Soon publications are scheduled from this telescope. According to the current plans three more LSTs, along with nine MSTs will be built and start operation in the coming several years in La Palma (CTA-webpage).

Similar to LST-1, the prototypes of the MST in Berlin, of the 9.7m diameter double mirror Schwarzschild-Couder telescope (SCT) in Arizona and the 4m prototype, double mirror ASTRI telescope in Sicily have been built and successfully commissioned, see (CTA-webpage) and the links therein. Both the SCT and the ASTRI telescope prototypes have detected signal from the Crab Nebula (Adams, et al., 2021),(Lombardi, et al., 2020). ASTRI has started construction of nine telescopes at the Teide Astronomical Observatory of the Instituto de Astrofisica de Canarias (IAC) in Tenerife, Canary Islands, Spain (Scuderi, Giuliani, Paresci, et al., 2022).

The Southern location of CTA is less than 10 km southeast of the European Southern Observatory's (ESO's) existing Paranal Observatory in the Atacama Desert in Chile.

In the current configuration, it is planned to build there 15+ MSTs, 37 SSTs and 3 LSTs. An expansion of the southern array by nine SSTs and a number of SCTs is under discussion.

One of the advances of the CTA telescopes can be considered the wide FoV of its telescopes. The systematic study of the wide-FoV prime-focus telescopes began in 2005 with publication (Schliesser, Mirzoyan, 2005). Soon this has been expanded by the study of even wider FoV IACTs of more complex design, including two optical elements (Vassiliev, et al., 2007),(Mirzoyan, Andersen, 2009). The SCT and ASTRI telescopes followed the design described in (Vassiliev, et al., 2007).
CTA is planning to operate LST, MST, SCT and SST telescopes of ~4.5°, ~8° and ~10° apertures.

The technology of these novel, 4$^{th}$ generation telescopes has been refined practically everywhere. After saturating the gamma-ray detection efficiency of individual telescopes, CTA is pursuing the plan to use a large number of such telescopes to cover a very large area. For example, the large number of SSTs will ensure a collection area of ~4 km$^2$ and thus exclusively high sensitivity. CTA will use advanced light sensors, such as the developed for its own use classical PMTs with worldwide best parameters (Mirzoyan, et al., 2017) as well as the so-called SiPM. Following the FACT telescope, operating the first full-scale SiPM-based camera (Neise, et al., 2017), also SCT and ASTRI are using SiPM-based cameras, see (CTA-webpage) and the links therein.

Whether the LSTs and MSTs can also be equipped with SiPM-based cameras in the future will depend on the expediency, cost development of these sensors and the readout channels as well as the availability of integrated readout solutions, see, for example, (Hahn, et al., 2022).

## 7.2 TAIGA

A hybrid approach has chosen the TAIGA pilot instrument in Siberia to access the gamma-ray energy range from tens to hundreds of TeV. Currently it includes 120 HiSCORE stations (up scaled and improved version of the AEROBICC detectors) deployed over an area of 1km$^2$ (Gress, et al., 2017) three 9.6° wide FoV, 4 m class IACTs (Budnev, et al., 2021) and other Cherenkov light and particle detectors. The 4$^{th}$ IACT is installed and is under completion. TAIGA plans to combine the timing and imaging EAS detection techniques in a novel "hybrid stereo" mode: the core position, direction of incidence and energy of a given shower are obtained from HiSCORE, while the largely spaced IACTs will allow measurement of the type (gamma or hadron) and also the energy of the primary. One of TAIGA's challenges was operating the telescopes in temperatures significantly below freezing, frequently reaching (-35)°C. This was solved by thermally isolating the imaging cameras, blowing warm air on the mirror segments, warming up the cable bundles, etc. Detection of the Crab Nebula signal by the first TAIGA IACT proved the successful construction (Blank, et al., 2023). TAIGA plans to install the telescopes at a distance of ~(600-800)m from each other. The 4 IACTs with HiSCORE form a sensitive detector with a detection area of over 1 km$^2$. The TAIGA approach promises to offer a cost-effective solution for building a highly sensitive, very large area detector.

## 7.3 LHAASO

LHASSO is a novel, multi-component, largest-ever built cosmic and gamma-ray detector. It is located in Daocheng, Sichuan Province of China, at an altitude of 4410m a.s.l. It aims to measure the spectrum, composition and anisotropy of cosmic rays in the energy range $\geq 10^{12}$ eV and gamma rays in the energy range $10^{11}$–$10^{15}$ eV. LHAASO can simultaneously measure different components of a shower such as electrons, muons, Cherenkov and fluorescence light. One of the components of LHAASO is the Wide Field-of-view Cherenkov Telescope Array (WFCTA), which consists of eighteen telescopes (Aharonian, et al., 2021). Each telescope consists of a segmented spherical mirror of about 5 m² with a SiPM-based camera installed in its focal plane. The imaging camera has a field of view (FoV) of 16°x16°. The telescopes are on trailers, so their location can be easily changed. In 2021, LHAASO made an important discovery; a dozen so-called PeVatron sources were identified (Cao, et al., 2021). Subsequently, it published a catalog of 90 sources with an extension below 2° and a detection significance > 5σ. Among these were 32 new sources (Cao, et al., 2023). Recently, they reported the discovery of an extremely intense signal (over 64000 photons with energy > 0.2 TeV) from the relatively nearby GRB 221009A, the brightest and most energetic gamma-ray burst ever recorded, observed from the initial moment and the transition to the afterglow (LHAASO Collaboration, 2023).

# CONCLUSIONS

It is impressive to follow the progress that has been made from the time of the first, tiny 25 cm diameter telescope used by Galbraith and Jelley in 1953, to the present day. After almost 70 years, Veritas, MAGIC, H.E.S.S. and now the LST/CTA telescopes can measure a gamma-ray signal from the Crab Nebula in less than a minute. Unexpectedly and surprisingly, Cocconi's 1959 speculative prediction came true. Today we operate instruments with a resolution of (0.05-0.1)°, which allow us to detect a gamma-ray signal from the Crab Nebula with a signal-to-noise ratio of ≥ 300:1 for energies ≥ 100 GeV. The CTA instrument, which will be completed in the near future, will further improve this signal-to-noise ratio. One should expect that breakthrough results in the fields of cosmic rays, multi-wavelength and multi-messenger astrophysics and cosmology will become available within the next ~10-15 years.